\documentclass[preprint,amsmath,amssymb,amsfonts,aps,prab,12pt,nofootinbib,obeyspaces,floatfix,superscriptaddress,showkeys]{revtex4-2}

\usepackage{graphicx}
\usepackage{dcolumn}
\usepackage{bm}
\usepackage{hyperref}

\usepackage{todonotes}
\usepackage{upgreek}
\usepackage{physics}
\usepackage{algpseudocode}
\newcommand{\Li}[2]{\text{Li}_{#1}\left({#2}\right)}
\usepackage{siunitx}
\begin{document}

\preprint{APS/123-QED}

\title{
Analytic descriptions of soft-edge quadrupoles \\for beamline design in multi-particle simulations
}

\author{A. Mahon}
\email{amahon@triumf.ca}
\affiliation{Department of Physics and Astronomy, University of Victoria}
\affiliation{TRIUMF, 4004 Wesbrook Mall, Vancouver, BC, Canada V6T 2A3}

\author{L. Miller}
\affiliation{TRIUMF, 4004 Wesbrook Mall, Vancouver, BC, Canada V6T 2A3}

\author{E. Cline}
\affiliation{Laboratory for Nuclear Science, Massachusetts Institute of Technology, Cambridge, MA 02139, USA}
\affiliation{Center for Frontiers in Nuclear Science, Stony Brook University, Stony Brook, NY 11790, USA}

\author{T. Planche} 
\affiliation{Department of Physics and Astronomy, University of Victoria}
\affiliation{TRIUMF, 4004 Wesbrook Mall, Vancouver, BC, Canada V6T 2A3}



\date{\today}

\begin{abstract}
All physical quadrupoles have a fringe field falloff, giving rise to intrinsic higher order aberrations. It is especially important to consider these aberrations when designing beamlines with strong, longitudinally cramped optics used to capture and transport large emittance beams. This paper presents relatively simple analytical formulae which describe the fields of quadrupoles including their higher order aberrations for implementation in beamline design.  We review the analytic model for quadrupoles described by a sech$^2$ strength function and subsequently expand it to quadrupoles whose strength function instead follows a tanh function, adding the effective length of the quadrupole as a free parameter. The implementation of these analytic field descriptions in envelope code {\sc transoptr} and multi-particle codes GEANT4 and FLUKA is presented using the optics design of the DarkLight experiment at TRIUMF as example. 

\end{abstract}

\maketitle


\section{\label{sec:intro}Introduction:}

When designing the optics of a beamline, it is common to exclude the fringe fields of optical elements, opting for the so-called ``hard-edge" approximation as a simple starting point. 
In some cases this is the logical order of events, as the elements are not yet designed and the precise shape of their fringe fields are not known. 
This approach is sufficient if the effect of the higher order aberrations can be neglected.

Quadrupoles have a finite length; even with perfect quadrupolar symmetry they give rise to intrinsic, irreducible higher order aberrations.
The lowest order intrinsic aberration, the cubic component, is independent of the fringe field shape~\cite{baartman2015quadrupoles}, and scales inversely with effective length ($L$) and the square of the focal length ($f^2$) of the quadrupole~\cite{baartman2018lowenergy}.
The smaller $f$ and $L$, the more significant the contribution of the intrinsic aberrations become, to a point where they can no longer be ignored.

This situation can arise in transport lines where secondary beams are produced by the interaction of high‑energy particles with a fixed target~\cite{Battaglieri:2023gvd}. 
To capture and transport the resulting large emittance beams, one tends to use strong quadrupoles (small $f$) placed in rapid succession (short $L$).
Take for example the optics of secondary muon beamlines~\cite{stratakis2019commissioning, cline2022characterization,TRI-BN-18-11},
which are tasked with transporting the desired beam within the constraints of the machine aperture. 
The field of these quadrupoles may also no longer have the ``flat top" shape as required for the hard-edge approximation~\cite{matsuda1972third}. 
Decelerating a beam, as is done for studying surface properties in condensed matter physics~\cite{morris2013beta} or in the case of antimatter studies, e.g.~measuring the free fall of antihydrogen~\cite{husson2021pulsed}, also results in transversely large beams requiring transport. 
Finally, any beamline or experiment needing to transport a highly scattered beam after collision with a target will also encounter these conditions.
An increasingly common application which introduces scattering to the beam is in medical physics. Depending on the nature of the medical accelerator the beam energy may not be so easily varied. 
A method used to attenuate such a beam is to pass it through a degrader, which reduces the energy but results in scattering, increasing the beam emittance~\cite{arjomandy2019aapm}.

When designing such beamlines it is essential to start not from a hard-edge approximation, but instead from a simple analytic ``soft-edge" model that takes into account the intrinsic aberrations from the outset. 
This paper builds on existing analytic 3-dimensional quadrupole models \cite{Baartman2012shapes,muratori2015analytical}, which are exact solutions to Maxwell's equations. We propose a simple field description with the minimum number of parameters required to independently adjust $L$ and $f$, thereby accurately representing the lowest order intrinsic aberrations. This approach is then demonstrated through its application in existing multi-particle tracking codes. 
Section~\ref{sec:analytic} reviews the analytic description of quadrupole fields with a $\sech^2$ strength function~\cite{Baartman2012shapes}, and generalizes this to a $\tanh$ strength function, providing one additional degree of freedom. An example implementation of these field descriptions is then presented in Section~\ref{sec:implementation} using the optics design of the DarkLight experiment at the TRIUMF electron linear accelerator (e-Linac)~\cite{cline2022searching}. 

In the DarkLight experiment, an electron beam is scattered off a Tantalum target; this results in a large emittance beam that must be recaptured and transported over a short distance to the beam dump, demonstrating the case of both a highly scattered beam and longitudinally cramped optics.
Each of the field descriptions presented in Section~\ref{sec:analytic} are used for the two types of quadrupoles present in the design; short electromagnetic quadrupoles from Buckley Systems, and longer permanent magnet quadrupoles from SABR Enterprises. 
This optics design is further used to present a comparison of the beam transport when only considering linear field components in the beam optics code {\sc transoptr}~\cite{heighway1981transoptr,baartman2016transoptr,transoptr_manual} 
to that of multi-particle codes GEANT4~\cite{allison2006GEANT4, allison2016recent,agostinelli2003GEANT4} and FLUKA~\cite{FLUKA0,FLUKA1, FLUKA2}, which include the higher order effects from the complete analytic descriptions.

\section{\label{sec:analytic}Analytic models}

This section details the analytic field descriptions of quadrupole magnets, whose fields focus the beam in one transverse direction and de-focus in the other. For the subsequent derivations and examples we adopt a right hand system of coordinates in which $x$ denotes the transverse horizontal direction, $y$ the transverse vertical direction, and $z$ the longitudinal direction along which the beam travels.

Note that in the thin lens approximation, the focal length $f$ of a quadrupole is inversely proportional to its integrated strength $K$. For the electrostatic case, $K$ is in units of electric field whereas for the magnetic case it is in units of magnetic field. From this point on we will refer to integrated strength in place of focal length.

\subsection{\label{sec:sech}Sech$^2$ Quadrupole}

Baartman \cite{Baartman2012shapes} describes the analytic expressions that can be used to model certain quadrupoles using a sech$^2$ strength function. This subsection will briefly review the methodology for introductory purposes.\\

Derevjankin~\cite{derevjankin1972representation,vasil1978spatial} proposed a formula that gives the scalar potential for a quadrupole with arbitrary on-axis field gradient profile $k(z)$:
\begin{equation}
    \label{eq:derevjankin}
    V(x,y,z) = -\Re{\int^{z+ix}_{z+iy}\int k(\zeta) \dd \zeta \dd t}\,.
\end{equation}
Baartman used this formula to obtain an explicit expression for the potential from a quadrupole with strength function $k(z) = \frac{K}{2}\text{sech}^2z$, where $K$ is the integrated strength of the element. The equation given in~\cite[Eq.\ (9)]{Baartman2012shapes} can alternatively be written as:
\begin{equation}
    \label{eq:sech2V}
    V(x,y,z) = -\frac{K}{4}  \log \left(\frac{\cos2x+\cosh2z}{\cos2y+\cosh2z}\right)\,.
\end{equation}

One can verify that, as expected, $\nabla^2V=0$. The explicit expressions for $\vec{F}=\nabla V$ given in \cite[Eq.\ (10--12)]{Baartman2012shapes} does not need to be reproduced here.\\

The electric field distribution from an electrostatic quadrupole is given by:
\begin{equation}
    \label{eq:FE}
    \vec{\mathcal{E}} = -\vec{F}\,,
\end{equation}
where $K$ is in units of electric field.\\

The same formula can be used to derive the magnetic field distribution from a magnetic quadrupole after rotating the scalar potential by 45 degrees:
\begin{equation}
    \vec{B}(x,y,z) = -\nabla V\left(\frac{x-y}{\sqrt{2}},\,\frac{x+y}{\sqrt{2}},\,z\right)\,,
\end{equation}
provided that $K$ is now units of magnetic field.
The application of the chain rule, and the fact that $\nabla \left(\frac{x-y}{\sqrt{2}},\,\frac{x+y}{\sqrt{2}},\,z\right)$ is the matrix of rotation around $z$ by an angle of 45 degrees, leads to:
\begin{equation}
    \label{eq:FB}
    \vec{B}(x,y,z) = -\vec{F}\left(\frac{x-y}{\sqrt{2}},\,\frac{x+y}{\sqrt{2}},\,z\right)\cdot    \left(
    \begin{array}{ccc}
            \frac{1}{\sqrt{2}}  & -\frac{1}{\sqrt{2}} & 0 \\
            \frac{1}{\sqrt{2}} & \frac{1}{\sqrt{2}} & 0 \\
            0                   & 0                  & 1 \\
        \end{array}
    \right)\,.
\end{equation}

\subsection{\label{sec:tanh}Tanh Quadrupole}

\begin{figure}[h!]
    \centering
    \includegraphics[width=0.5\linewidth]{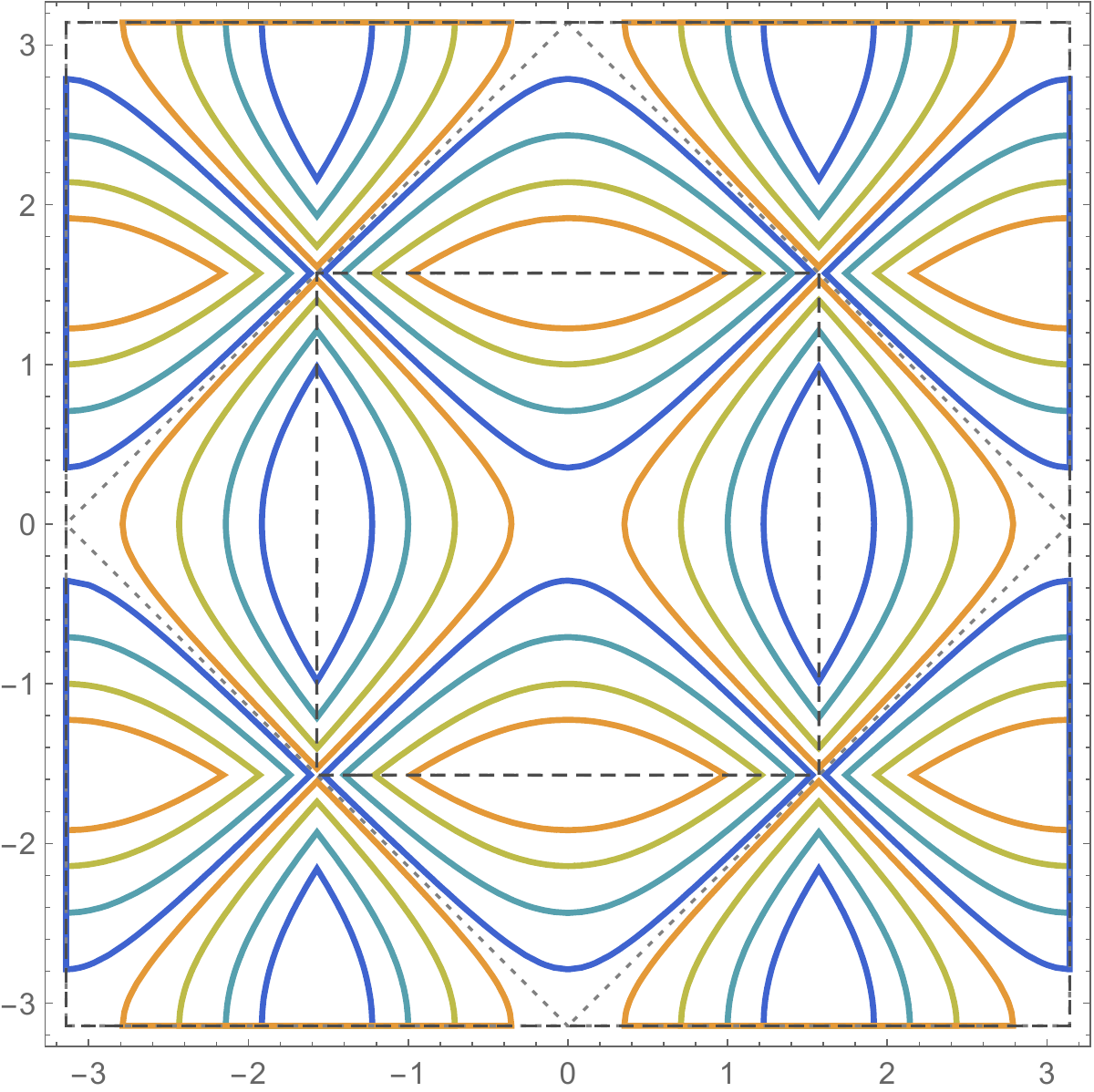}
    \caption{Equipotentials of Eq.~\ref{eq:pot} in the z = 0 plane.
    \label{fig:eqp}
    The vertical contours correspond to $\text{V}=3V_0/2$ (blue), $\text{V}=V_0$ (teal), $\text{V}=V_0/2$ (green), and $\text{V}=V_0/8$ (orange). The horizontal contours correspond to $\text{V}=-V_0/8$ (blue), $\text{V}=-V_0/2$ (teal), $\text{V}=-V_0$ (green), and $\text{V}=-3V_0/2$ (orange). The dashed square represents the field limit as discussed in Section~\ref{sec:tanhlim}. The dotted line gives an approximation to the e-Linac beam pipe diameter.
    }
    
\end{figure}

For quadrupoles that do not have a $\text{sech}^2$-like strength function, one can try to use the more general Enge function~\cite{enge1964effect} to fit the field fall-off on either side of the magnet. Here we consider only the case of a quadrupole with identical entrance and exit edges, separated by a distance $L$, and with all Enge coefficients equal to 0 except for $c_1 = 2/\lambda$, where lambda denotes the fringe field extent parameter. The strength function of our quadrupole becomes:

\begin{eqnarray}
      k(z) = \frac{K}{L} \left( \frac{1}{1+e^{2\left(\frac{z-L/2}{\lambda}\right)}}-\frac{1}{1+e^{2\left(\frac{z+L/2}{\lambda}\right)}}\right)\,.
\end{eqnarray}
Note that $\int_{-\infty}^\infty k(z)\dd z = K$, the integrated strength of our quadrupole.
To keep the math nice and tidy we choose hereunder our unit of length so that $\lambda = 1$.
Using the identity $\tanh x = 1-\frac{2}{1+e^{2x}}$ we re-write:
\begin{equation}
    k(z) = \frac{K}{2 L} \left( \tanh(z+L/2)-\tanh(z-L/2)\right)\,.
    \label{eq:tanh}
\end{equation}
Using Eq.~\ref{eq:derevjankin} leads to a potential $V$ =
\begin{equation}
    \frac{K}{4L}\Re{ \Li{2}{-e^{L+2(ix+z)}}\!-\!\Li{2}{-e^{L+2(iy+z)}}\!-\!\Li{2}{-e^{-L+2(ix+z)}}\!+\!\Li{2}{-e^{-L+2(iy+z)}} \,},
    \label{eq:pot}
\end{equation} 
where $\Li{2}{t}=-\int_0^t{\frac{\ln{(1-z)}}{z} \dd z}$ is the polylogarithm function of order 2. The equipotentials for this expression are drawn in Fig.~\ref{fig:eqp}, using $V_0 = V(1,0,0) = \frac{K}{4L}$ as scaling potential. 

Admittedly, this expression would not be very useful except for the fact that its gradient depends only on conventional trigonometric functions:
\begin{equation}
    \begin{aligned}
        F_x &= \frac{K}{2 L} \arctan(\frac{\sinh L \sin 2 x}{\cosh L \cos 2 x+\cosh 2 z})\,,\\
        F_y &=-\frac{K}{2 L} \arctan(\frac{\sinh L \sin 2 y}{\cosh L \cos 2 y+\cosh 2 z})\,,\\
        F_z &= \frac{K}{4 L} \log(\frac{(\cosh (L-2 z)+\cos 2 x) (\cosh (L+2z)+\cos 2 y)}{(\cosh (L+2 z)+\cos 2 x) (\cosh (L-2z)+\cos 2 y)})\,.
    \end{aligned}   
    \label{eq:tanh_fields}
\end{equation}

While these are the explicit expressions, it is in practice easier to define these functions more logically:

\begin{equation}
    \begin{aligned}
        R(x,y,z,L) &= \arctan \left( \dfrac{\sinh L \sin 2x}{\cosh L \cos 2x + \cosh 2z} \right)\,, \\
        S(x,y,z,l) &= \frac{1}{2}\log \left( (\cosh(L-2z)+\cos 2x)(\cosh(L+2z)+\cos 2y)\right)\,.
    \end{aligned}
\end{equation}

Combining the above functions we can construct our final expression for the fields:

\begin{equation}
        \vec{F} = \dfrac{K}{2L}\left[ \begin{array}{c}
            R(x,y,z,L) \\
            -R(y,x,z,L)\\
             S(x,y,z,L)-S(y,x,z,L)\\
        \end{array}
        \right] \,.
\end{equation}

Here again one can use Eq.~\ref{eq:FE} or Eq.~\ref{eq:FB} to obtain the corresponding electric or magnetic field distribution. A reminder that these equations are still operating under the choice of unit length such that $\lambda$ = 1.
We have included a pseudo-code in Appendix~\ref{sec:appendix} with this simplified logic and derive the respective electric and magnetic field distributions for ease of implementation in analysis codes. \\ 

The two descriptions detailed above are ultimately not so different from one another. The sech$^2$ fit presents the option with the lowest degrees of freedom, whereas the tanh description is simply a more general case with one extra fitting parameter, that being the quadrupole length $L$. It can be shown that in the limit of short lengths, the tanh function simplifies back to the sech function:

\begin{equation}
    \lim_{L\to 0} \frac{K}{2 L} \left( \tanh(z+L/2)-\tanh(z-L/2)\right)\ = \frac{K}{2}\text{sech}^2z \,.
\end{equation}

\subsection{\label{sec:tanhlim}Boundary of validity of field model}

The potential derived in the previous section is continuous, periodic and defined over all space. However, the fields that we derive from it in Eq.~\ref{eq:tanh_fields} present a discontinuity in the transverse plane illustrated by the dashed line in Fig.~\ref{fig:eqp}. Therefore, this soft-edge model only represents a realistic quadrupole within this boundary.
The distance from the center of the quadrupole to this point along one axis depends on the fringe field extent parameter $\lambda$. As we wish to operate within this boundary, let us now derive at how many $\lambda$ away the model is valid.

It is important to note that the arctan function used in Eq.~\ref{eq:tanh_fields} is the four quadrant inverse tangent function, which is defined over the range (-$\pi$, $\pi$]. 
It remains to evaluate where the arctan terms of the transverse field equations reach this boundary, which occurs when the argument is equal to zero.
As $L$ $>$ 0, only the $\sin2x$ term contributes to the argument vanishing, which in turn leads to the solution being multiples of $\pi$/2. Therefore, the limit of a quadrupole's region of validity in our model is, for the electrostatic case:

\begin{equation}
    x = \frac{\lambda \pi}{2}\,.
\end{equation}

This describes a square good-field region with side of $\lambda \pi$ for the electrostatic quadrupole, which is drawn over the equipotential lines in Fig.~\ref{fig:eqp}. It follows for a magnetic quadrupole that the distance from the center of the quadrupole to the boundary along the x axis is:

\begin{equation}
    x = \frac{\sqrt{2}\lambda \pi}{2}
\end{equation}

\noindent{which draws a diamond. So long as this limit is taken into account, one can adjust the aperture and fringe field extent of the optical element being designed to ensure the field description will hold for the entirety of the magnet aperture.}

\subsection{\label{beyond_tanh} Beyond tanh}

The analytic descriptions presented above are not meant to generalize the fringe fields of all optical elements. This approach falls short, for example, for elements with very sharp fringe field fall off or with uneven entrance and exit edges. 
This is in part due to the tanh function being an exactly odd function. In real quadrupoles, the gradient begins to increase in a more gradual fashion than the tanh function, and about midway switches to approach the final value more abruptly. 
However, it has been shown that the leading third order aberrations of quadrupoles are insensitive to this asymmetry \cite{baartman2015quadrupoles}, so the tanh approximation still works better than one might assume.
The Enge function \cite{enge1964effect} could be used to introduce further degrees of freedom for these cases, but we lose the simplicity promised by this method. This brings up the question of whether it is more important to precisely fit the shape of the edge, or to work with a general shape of the fringe fields. In this case one has two options:

\begin{enumerate}
    \item Determine to what level the precision of your elements is important for your application. Using an analytic description of the fields that assumes the symmetry of entrance and exit edges may not be perfectly exact, but it will never yield nonphysical results. 
    \item Where possible you can reverse this approach, and intentionally choose the shape of the poles or coils of the optical elements such that their fringe fields will be accurately described by the aforementioned analytic descriptions. In the case of magnets that are already built, one could shim the edges of the yoke to achieve the same result. 
\end{enumerate}

We will now demonstrate the implementation of the analytic formulas derived above in both linear and multi-particle codes, and highlight the effect of field non-linearities. 

\section{\label{sec:implementation}Implementation}

\begin{figure}[ht]
    \centering
    \includegraphics[width = 0.56\textwidth]{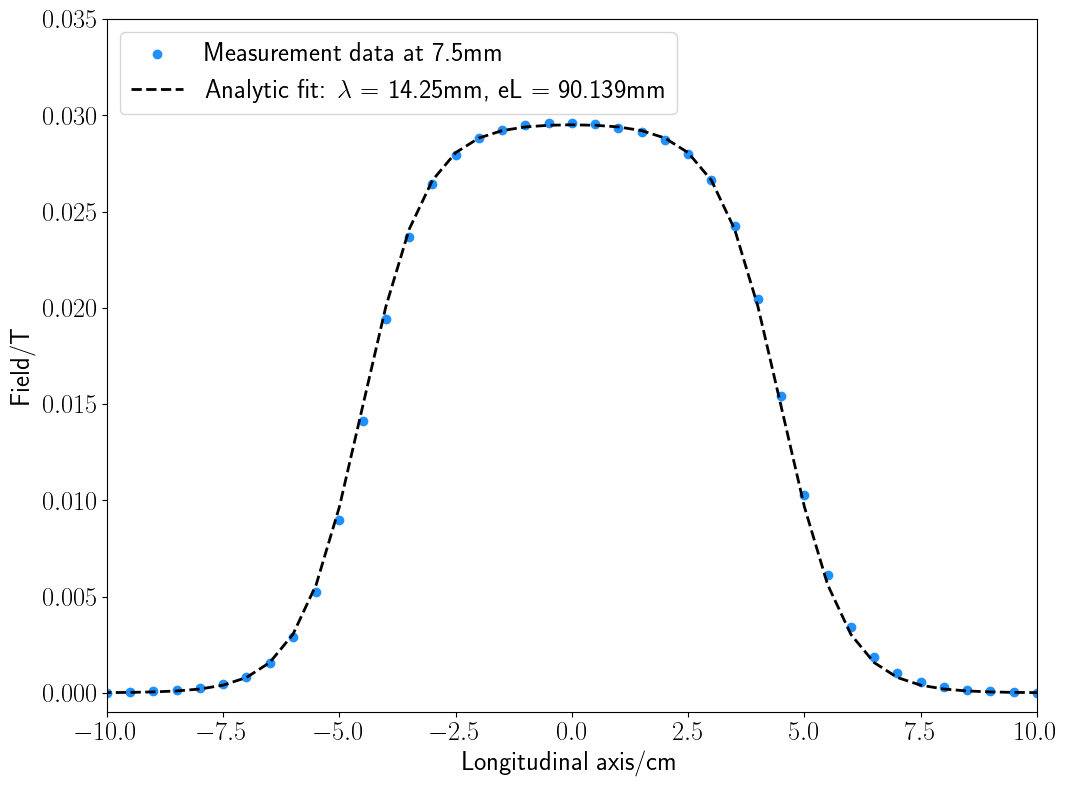}
    \caption[short]{
    Field measured at 7.5\,mm off axis for a SABR permanent magnet quadrupole parameterized using the equations derived in Section~\ref{sec:tanh}. Fit parameters: $K = 0.356\,\text{T}, \lambda = 14.25\,\text{mm}, L = 90.14\,\rm{mm}$.}
    \label{fig:magnet_fitting}
\end{figure}

    The optics design used in the subsequent examples comes from the downstream optics of the DarkLight experiment at the TRIUMF electron linear accelerator~\cite{cline2022searching},  where a 31\,MeV electron beam strikes a 1\,\SI{}{\micro\meter} Ta target. After this interaction the scattered beam is transported to the beam dump through the following optics: a triplet of permanent magnet quadrupoles, whose fields are most accurately modeled using the tanh description (fit parameters shown in Fig.~\ref{fig:magnet_fitting})    
    and an electromagnetic doublet, best modeled by the sech$^2$ description (fit parameter $\lambda$ = 33.1\,mm \cite{Baartman2013buckley}).
    Further magnet specifications are listed in Tab.~\ref{tab:PMQs} and  Tab.~\ref{tab:EMQs} respectively, and details of the magnetic measurements can be found in \cite{TRI-BN-23-28} and \cite{TRI-BN-13-12}.
    The details of this optics design can be found in TRIUMF design note TRI-DN-23-17 \cite{TRI-DN-23-17}\footnote{Internal TRIUMF documents are available upon request.}, with a visual reconstruction provided in Fig.~\ref{fig:FLUKA_Beamline}.

A small note on the fit shown in Fig.~\ref{fig:magnet_fitting}; as the magnetic measurement for the permanent magnets was done off axis, it was not possible to fit the tanh strength function directly as it is only valid for the on axis gradient. Instead it was necessary to fit the fringe field parameter $\lambda$ using the complete 3D magnetic field description (see Appendix~\ref{sec:appendix} for implementation). This was required as the field on axis will not have the same shape as we approach the aperture of the magnet; in fact it begins to resemble a shape akin to the hard-edge step like function. \\

\begin{table}[ht]
\caption{Parameters of DarkLight permanent magnet quadrupoles}
\begin{ruledtabular}
\begin{tabular}{lclc}
\textrm{Manufacturer} & \textrm{SABR Enterprises, LLC} &
\textrm{Magnetic Material} & \textrm{SmCo$_{2:17}$} \\

\textrm{Integrated Strength} & $0.3\,\mathrm{T} \pm 5\%$ &
\textrm{Inner diameter} & $5.2\,\mathrm{cm}$ \\

\textrm{Nominal Gradient} & $3.33\,\mathrm{T/m}$ &
\textrm{Outer diameter} & $7.0\,\mathrm{cm}$ \\

\textrm{Length} & $9.3\,\mathrm{cm}$ &
\textrm{Number of Halbach segments} & $16$
\end{tabular}
\end{ruledtabular}
\label{tab:PMQs}
\end{table}

\begin{table}[ht]
\caption{Parameters of DarkLight electromagnetic quadrupoles}
\begin{ruledtabular}
\begin{tabular}{lclc}
\textrm{Manufacturer} & \textrm{Buckley Systems} &
\textrm{Aperture Diameter} & $52.00\,\mathrm{mm}$ \\

\textrm{Max Integrated Strength} & $0.7\,\mathrm{T}$ &
\textrm{Max Operating Current} & $39.0\,\mathrm{A}$
\end{tabular}
\end{ruledtabular}
\label{tab:EMQs}
\end{table}

    The tool used at TRIUMF to model beam envelope transport is the linear optics code {\sc transoptr}~\cite{heighway1981transoptr,baartman2016transoptr,transoptr_manual}.
    A unique feature of the comparisons presented below is that, rather than comparing a hard-edge approximation to a soft-edge, we are comparing a truncated (linear) soft-edge ({\sc transoptr}) to the full non-linear soft-edge in multi-particle codes (GEANT4 and FLUKA). Thus, any differences we see in the beam transport at large beam sizes can be attributed to the inclusion of higher order aberrations present in the codes which use the untruncated, full analytic description of the fields.

\subsection{\label{sec:initcon}Initial conditions}

The initial beam conditions given to {\sc transoptr} are obtained by taking the statistics of the distribution of the events from both GEANT4 and FLUKA at a location 1\,cm after the interaction target. 
Computing the covariance matrix (often called beam matrix or sigma matrix) of the particle distribution provides the necessary information regarding initial size ($x_{\rm rms}/y_{\rm rms}$) and angular spread ($x'_{\rm rms}/y'_{\rm rms}$) of the beam, in addition to the correlation between the two ($r_{12/34})$.
The initial conditions for all following example cases can be found in Tab.~\ref{tab:ics} and Tab.~\ref{tab:ics_flk}.

\begin{table}[h!]
\caption{Initial conditions and correlation values obtained from GEANT4 simulations for various target thicknesses with a 31 MeV incident electron beam.}
\begin{ruledtabular}
\begin{tabular}{lrrrrrr}
\textbf{Target} & $x_{\mathrm{rms}}~(\mathrm{mm})$ & $x'~(\mathrm{mrad})$ & $y_{\mathrm{rms}}~(\mathrm{mm})$ & $y'~(\mathrm{mrad})$ & $r_{12}$ & $r_{34}$ \\

no target 
& 0.327 
& 0.215 
& 0.670 
& 0.749 
& 0.0160 
& 0.00859 \\

$1~\SI{}{\micro\meter}$  
& 0.332 
& 5.658 
& 0.673 
& 5.64 
& 0.167 
& 0.0836 \\

$5~\SI{}{\micro\meter}$  
& 0.348 
& 11.9 
& 0.681 
& 12.0 
& 0.339 
& 0.175 \\

$10~\SI{}{\micro\meter}$  
& 0.363 
& 15.9 
& 0.689 
& 16.1 
& 0.435 
& 0.233
\end{tabular}
\end{ruledtabular}
\label{tab:ics}
\end{table}

\begin{table}[h!]
\caption{Initial conditions and correlation values obtained from FLUKA simulations for various target thicknesses with a 31 MeV incident electron beam.}
\begin{ruledtabular}
\begin{tabular}{lrrrrrr}
\textbf{Target} & $x_{\mathrm{rms}}~(\mathrm{mm})$ & $x'~(\mathrm{mrad})$ & $y_{\mathrm{rms}}~(\mathrm{mm})$ & $y'~(\mathrm{mrad})$ & $r_{12}$ & $r_{34}$ \\

$1~\SI{}{\micro\meter}$  
& 0.334 
& 5.51 
& 0.673 
& 5.54 
& 0.249 
& 0.124 \\

$5~\SI{}{\micro\meter}$  
& 0.368 
& 11.8 
& 0.690 
& 11.9 
& 0.479 
& 0.258 \\

$10~\SI{}{\micro\meter}$  
& 0.402 
& 16.0 
& 0.710 
& 16.4 
& 0.595 
& 0.343
\end{tabular}
\end{ruledtabular}
\label{tab:ics_flk}
\end{table}

\subsection{\label{sec:GEANT4} Comparison of {\sc transoptr}  envelope to GEANT4}

A GEANT4~\cite{allison2006GEANT4, allison2016recent,agostinelli2003GEANT4} model of the beamline was implemented. The simulation includes the magnetic field as described by the analytic expressions in the previous sections, the appropriate electron beam parameters, and the DarkLight target (nominally a 1\,\SI{}{\micro\meter} thick tantalum foil). 

The purpose of this simulation was two-fold: first, {\sc transoptr} does not model the scattering of the beam through the target material, and thus the GEANT4 simulation is used to determine the values of the beam matrix 
supplied to {\sc transoptr} to simulate the beam after scattering has occurred. Second, the GEANT4 results serve as a cross check and comparison to the {\sc transoptr} result. As GEANT4 is a higher order multi-particle code, it will take into account all non-linearities, whereas {\sc transoptr} is a purely linear model. We therefore expect larger discrepancies between the models for cases with greater scattering, as the beam will occupy a larger fraction of the magnet aperture where the non-linear effects are more significant. To demonstrate this, we have simulated the optics for several different target thicknesses.

To determine the input parameters for the beam matrix used in {\sc transoptr}, the position and angle of each particle as it exits the target are recorded from the GEANT4 simulation, and the statistical moments are calculated. The initial conditions for all example cases can be found in Tab.~\ref{tab:ics}. As {\sc transoptr} does not handle individual particle losses when tracking the beam envelope, particles in the GEANT4 simulation which depart the radial extent of any of the individual magnetic fields are removed. The RMS of the surviving particle distribution is calculated at 1\,cm intervals along the length of the simulation for comparison to the {\sc transoptr} envelopes.

\begin{figure}[h!]
    \centering
    \includegraphics[width=\linewidth]{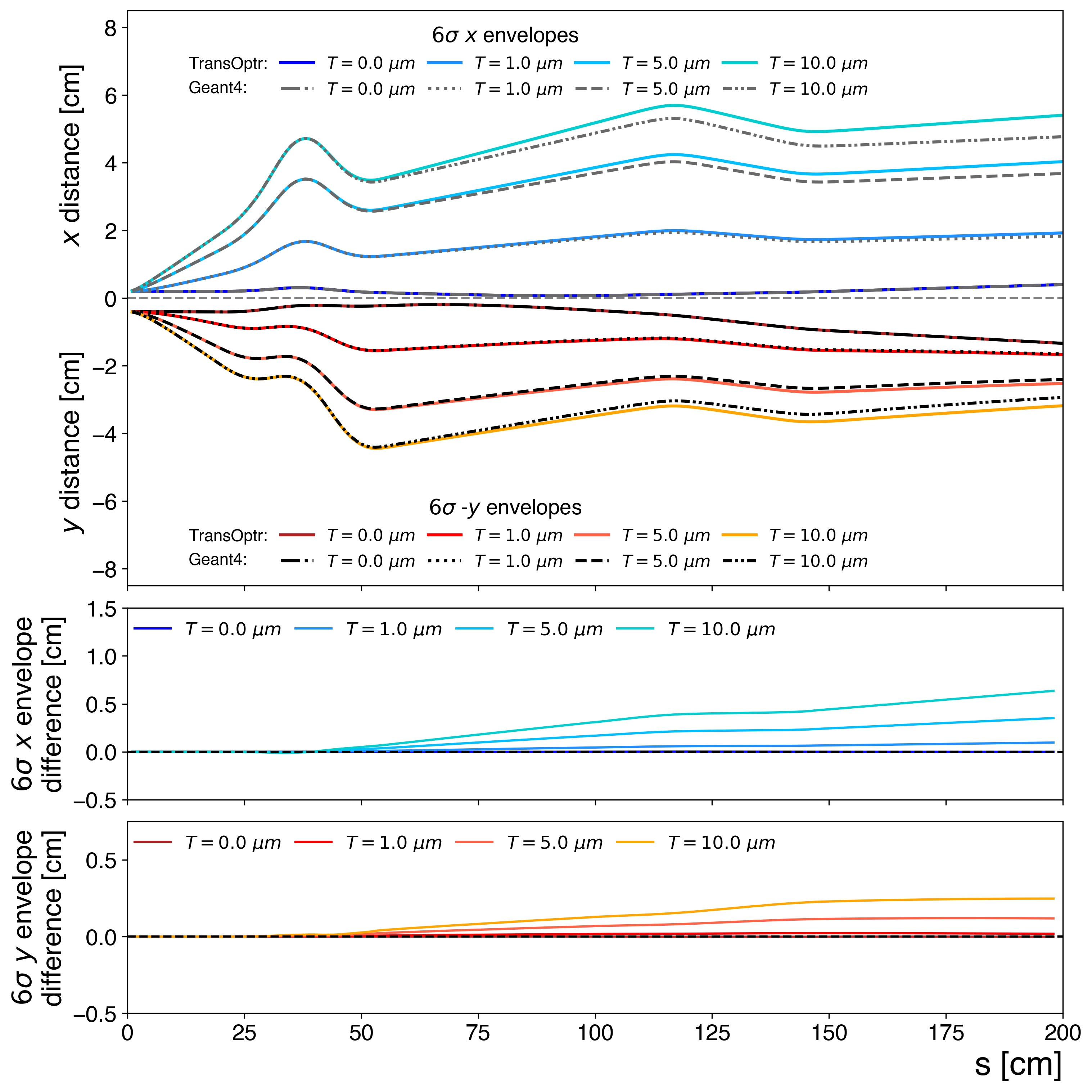}
    \caption{$6\sigma$ $x$ and $y$ beam envelopes and their corresponding differences from {\sc transoptr} and GEANT4 for multiple target thicknesses ($T$): no target, 1.0\,\SI{}{\micro\meter}, 5.0\,\SI{}{\micro\meter}, and 10.0\,\SI{}{\micro\meter}. Solid colored lines correspond to {\sc transoptr} outputs, dashed lines correspond to GEANT4 outputs.}
    \label{fig:scatterG4}
\end{figure}

The first relevant comparison is the case with no scattering.  The beam envelopes are expected to match exactly between {\sc transoptr} and GEANT4, as with no scattering the beam remains narrow and well contained, thus it will see none of the non-linearities at large apertures. The result of this comparison can be seen in Fig.~\ref{fig:scatterG4}. As expected, the two envelopes match exactly.
The beam envelopes for target thicknesses of 1.0\,\SI{}{\micro\meter}, 5.0\,\SI{}{\micro\meter}, and 10.0\,\SI{}{\micro\meter} are also shown in Fig.~\ref{fig:scatterG4}. 
The discrepancies arising from the nonlinear effects can be seen when comparing these three cases of increasing target thicknesses. The $6\sigma$ beam envelopes show increasing divergence between {\sc transoptr} and GEANT4 as the target thickness gets larger, indicative of the non-linearities coming into effect in GEANT4 as the beam begins to occupy more of the magnet apertures from the increased scattering.

\begin{figure}[h!]
    \centering
    \includegraphics[width=0.49\linewidth]{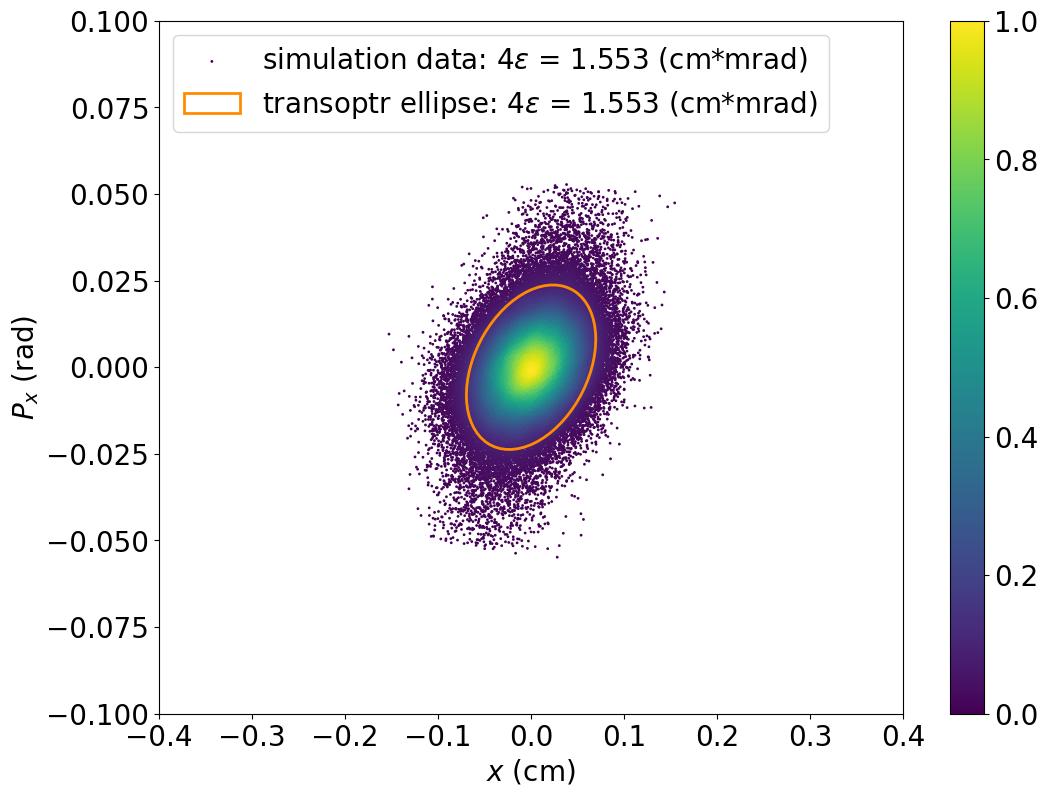}
    \includegraphics[width = 0.49\linewidth]{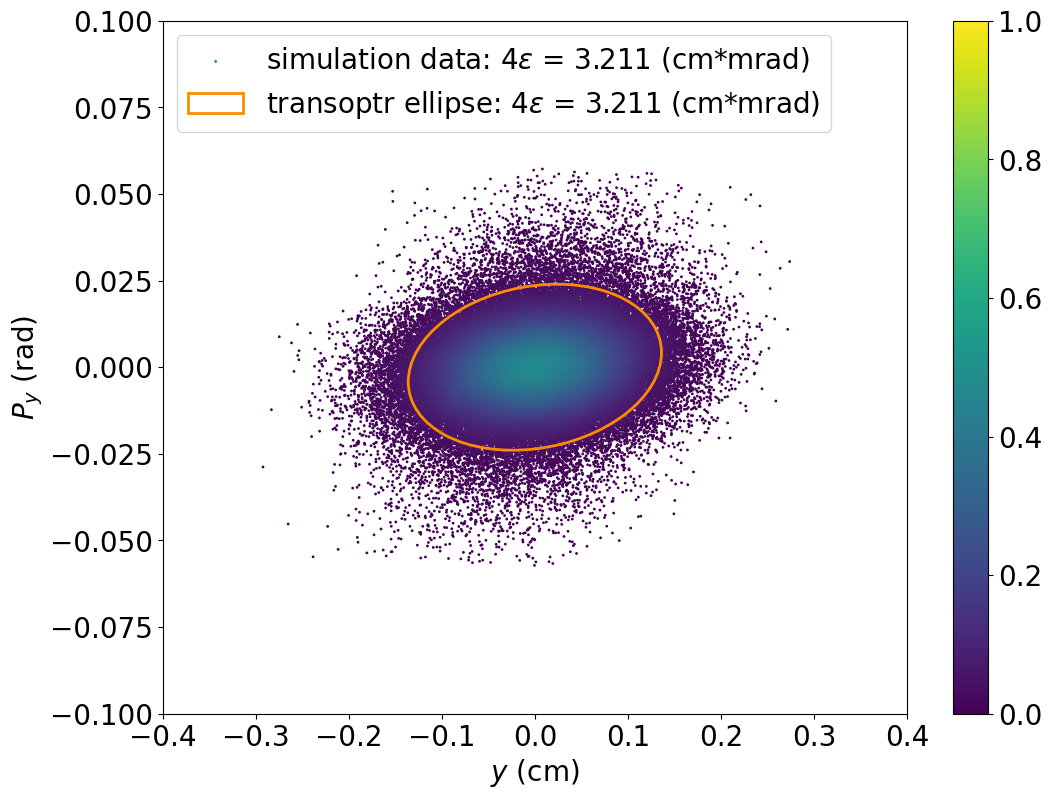}
    \includegraphics[width=0.49\linewidth]{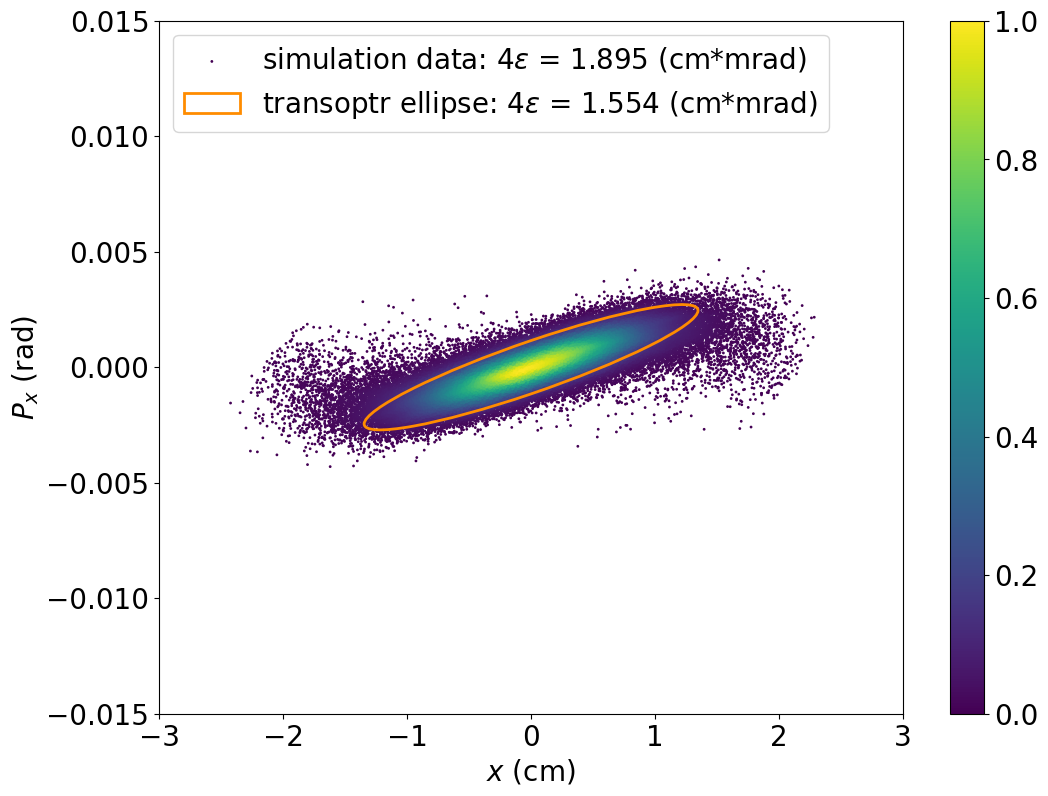}
    \includegraphics[width = 0.49\linewidth]{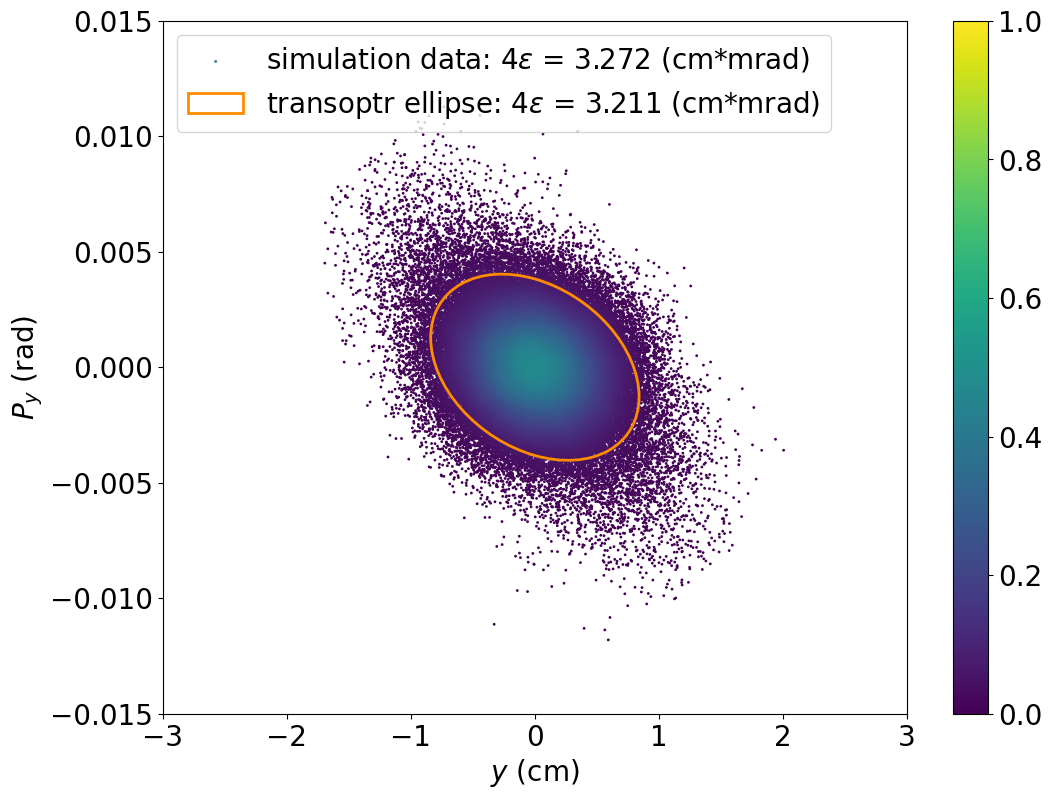}
    \caption{Phase space distributions of particles generated in GEANT4 compared to the phase space ellipse obtained from {\sc transoptr} for the 5\,\SI{}{\micro\meter} target run.
    Top left: $x$ phase space at start of simulation; Top right: $y$ phase space at start of simulation; Bottom left: $x$ phase space at the end of simulation; Bottom right: $y$ phase space at end of simulation. Colorbars represent the normalized density of particles.}
    \label{fig:g4_phasespace}
\end{figure}

The phase space distribution of particles along the beamline provides additional insight into the effect of the non-linearities. Figure~\ref{fig:g4_phasespace} shows the phase space distributions in both the $x$ and $y$ planes of the 31\,MeV electrons at the start and end of the GEANT4 simulation for the 5\,\SI{}{\micro\meter} target run. These are overlaid with the phase space ellipse obtained from the {\sc transoptr} simulation output. The area of this ellipse describes the emittance of the beam, a property used to quantify beam quality and size.
The typical phase space distribution for a diverging beam will draw an ellipse tilted to the right. The upper plots in Fig.~\ref{fig:g4_phasespace} demonstrate this shape nicely. 
Liouville's theorem \cite{bruck1972circular} states that under the action of conservative forces, the emittance of the beam is conserved. When a beam undergoes non-linear effects it will experience an increase in emittance, which can be visualized by a characteristic ``S" shape in the phase space distribution. This is seen in the lower two plots of Fig.~\ref{fig:g4_phasespace}, particularly in the x direction, once the beam has been transported through the optical elements. The emittance values are also displayed on the plot for comparison: the GEANT4 emittance increases in both the x and y planes, whereas the {\sc transoptr} emittance is conserved. 

\subsection{\label{sec:FLUKA}Comparison of {\sc transoptr} envelope to FLUKA}

\begin{figure}[h!]
    \centering
    \includegraphics[width=\linewidth]{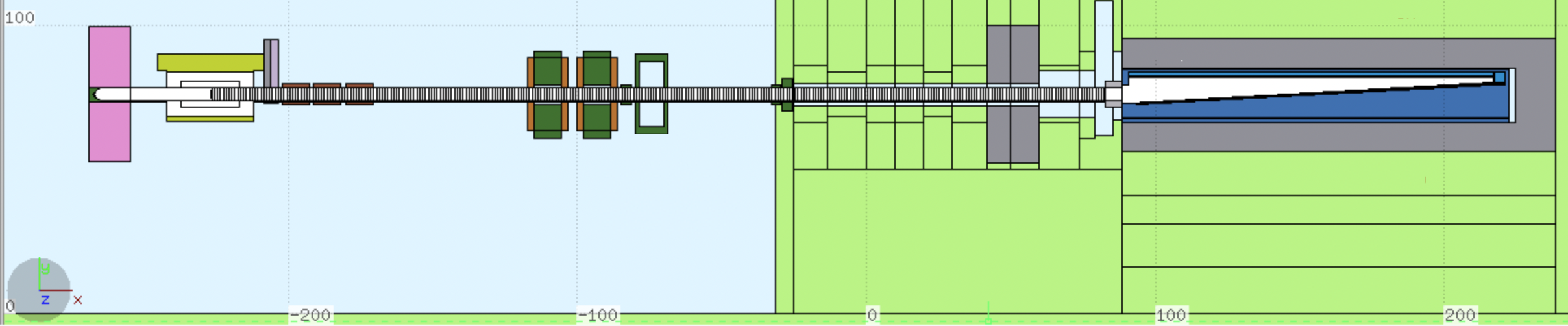}
    \caption{Cutaway view of the beamline as implemented in FLUKA and visualized in FLAIR. In this figure the beam travels from left to right. The beam pipe, situated 76 cm above the floor, from left to right, connects the final bending dipole of the e-Linac beamline, the DarkLight target chamber and shielding, three permanent quadrupoles, two electromagnetic quadrupoles, and a diagnostic box. The beam dump is shown in blue on the right, surrounded by lead shielding shown in gray, and concrete shown in green. The units on the axes in the figure are in cm.}
    \label{fig:FLUKA_Beamline}
\end{figure}

As a multi-particle code, FLUKA~\cite{FLUKA0,FLUKA1, FLUKA2} also models the scattering of beam particles onto the target given fixed energy, beam size, target material and thickness inputs.
FLUKA was used to simulate the interaction of a 31 MeV electron beam onto the DarkLight target.
The e-Linac beam dump geometry and input deck had been previously developed \cite{TRI-DN-13-29,TRI-DN-13-11}.
Modifications to the geometry to reflect the DarkLight experiment~\cite{cline2022searching}, in particular, the beam pipe and focusing quadrupoles, were implemented.
The beamline can be visualized in FLAIR~\cite{FLAIR}, as seen in Fig.~\ref{fig:FLUKA_Beamline}.

The magnetic field is described in a user-defined input routine MAGFLD \cite[section 13.2.12]{ferrari2005fluka} using the analytic formula described in Section~\ref{sec:analytic}, to ensure an identical treatment of the optics as in the case of GEANT4, cf. Section~\ref{sec:GEANT4}. 
The beam is generated from a gaussian distribution with an RMS width identical to that in {\sc transoptr}. The beam originates several centimeters upstream of the tantalum target, propagates through the target, and scatters. The beam particles are recorded in a USRBIN sensitive detector, designed to record particle fluence. A cut is once more applied to particles which depart the radial extent of any of the magnetic fields. The resulting beam distribution at the start of the FLUKA simulation is used to compute the initial conditions for the {\sc transoptr} comparison post target, with these values being recorded in  Tab.~\ref{tab:ics_flk}.

\begin{figure}[h!]
    \centering
    \includegraphics[width=\linewidth]{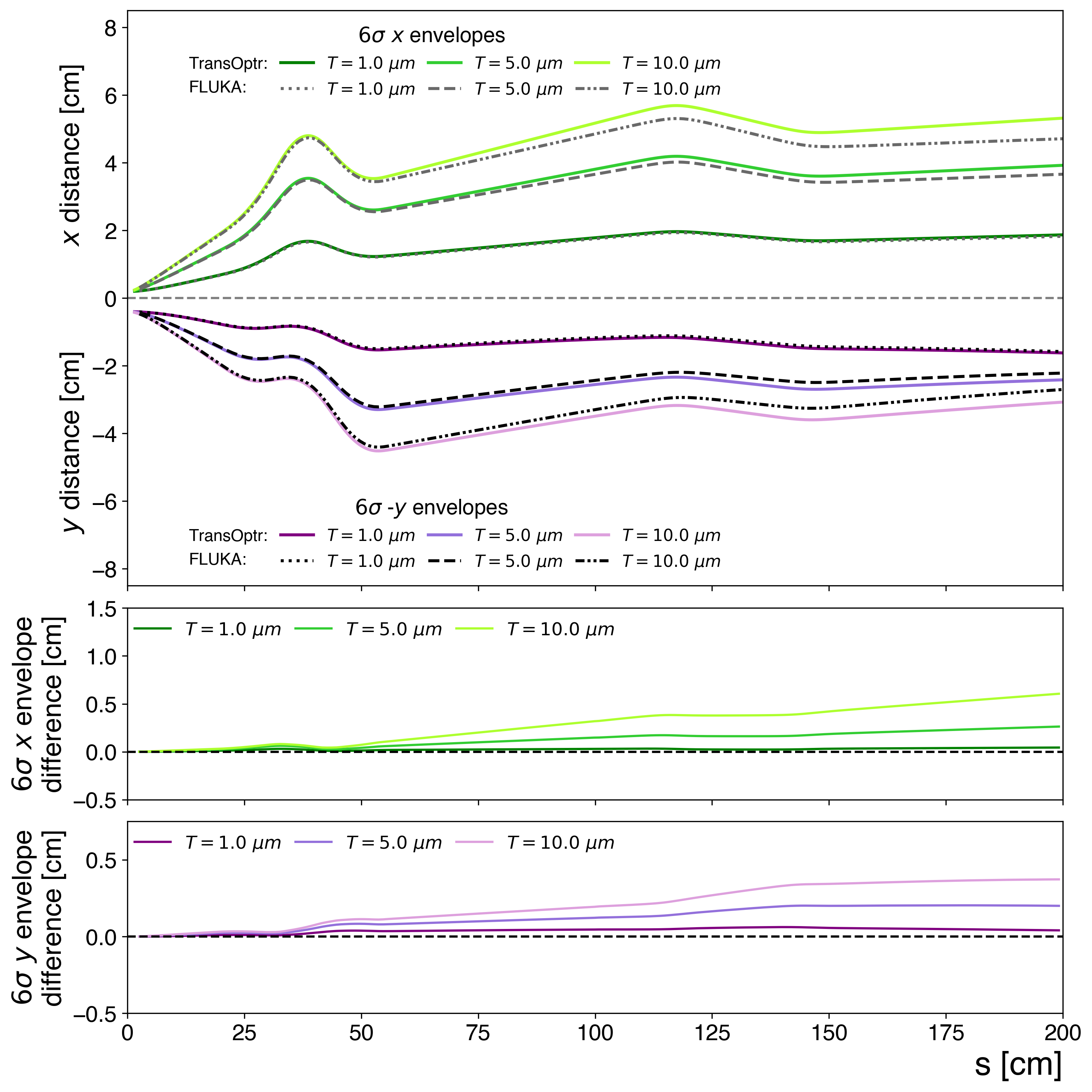}
    \caption{$6\sigma$ $x$ and $y$ beam envelopes and their corresponding ratios from {\sc transoptr} and FLUKA for multiple target thicknesses ($T$): no target, 1.0\,\SI{}{\micro\meter}, 5.0\,\SI{}{\micro\meter}, and 10.0\,\SI{}{\micro\meter}. Solid colored lines correspond to {\sc transoptr} outputs, dashed lines correspond to FLUKA outputs.}
    \label{fig:scatterFLK}
\end{figure}

\begin{figure}[h!]
    \centering
    \includegraphics[width=0.49\linewidth]{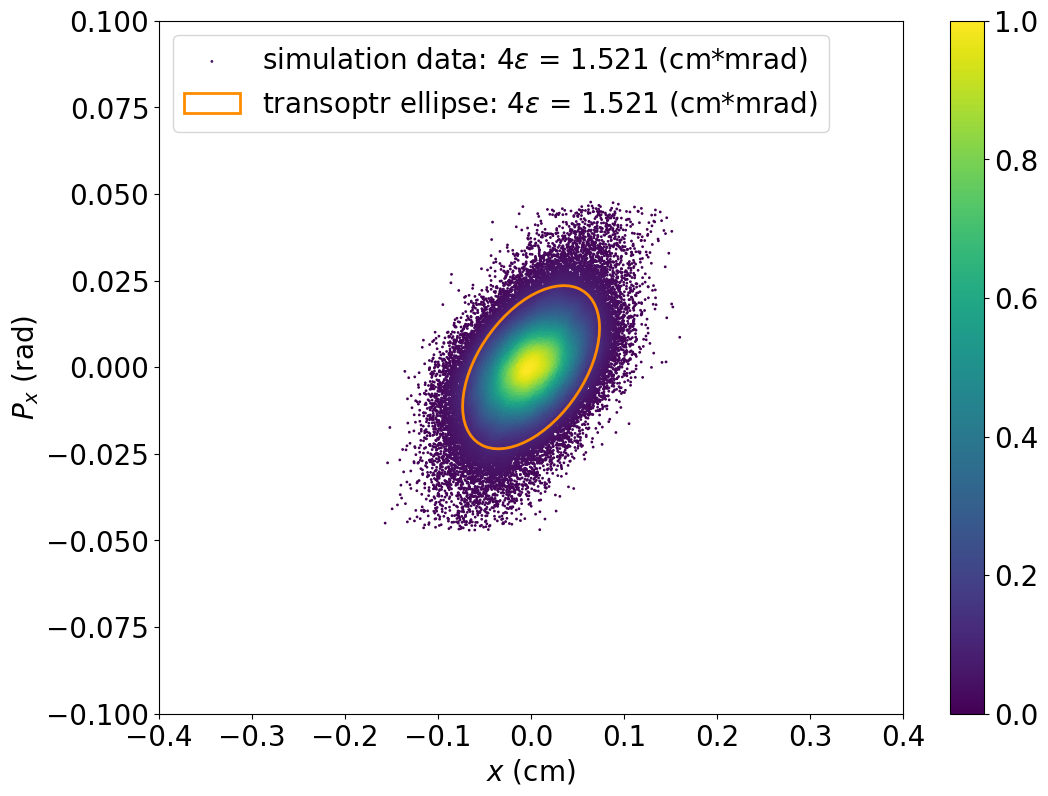}
    \includegraphics[width = 0.49\linewidth]{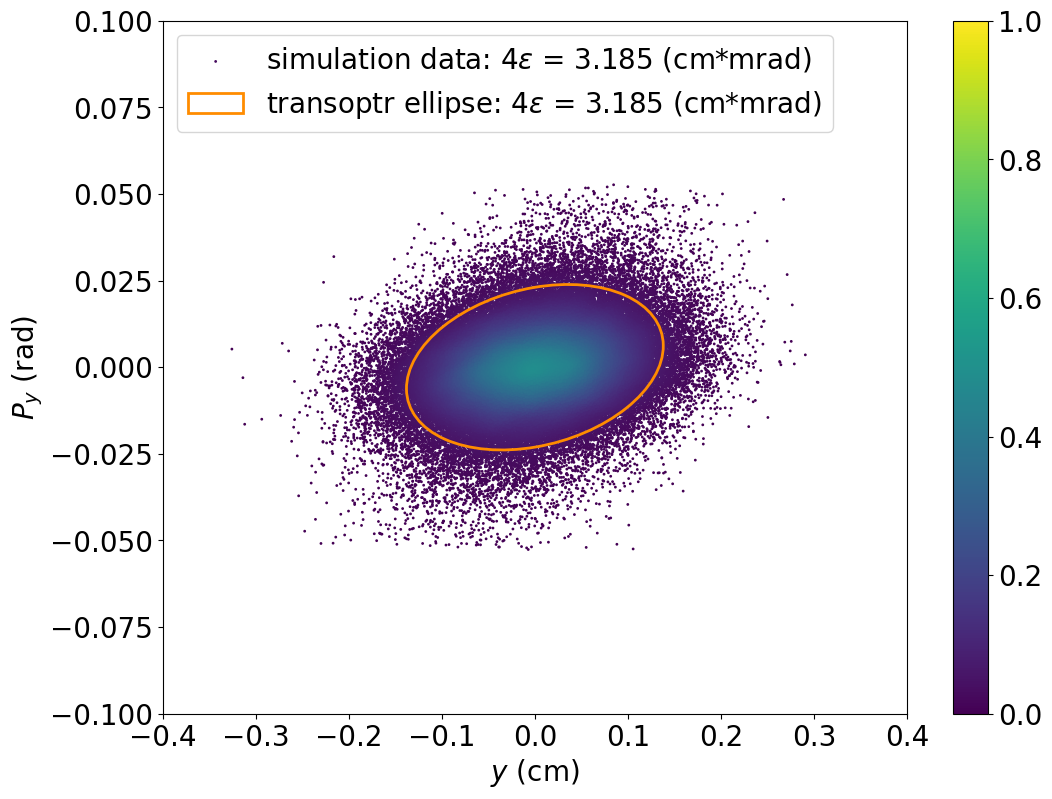}
    \includegraphics[width=0.49\linewidth]{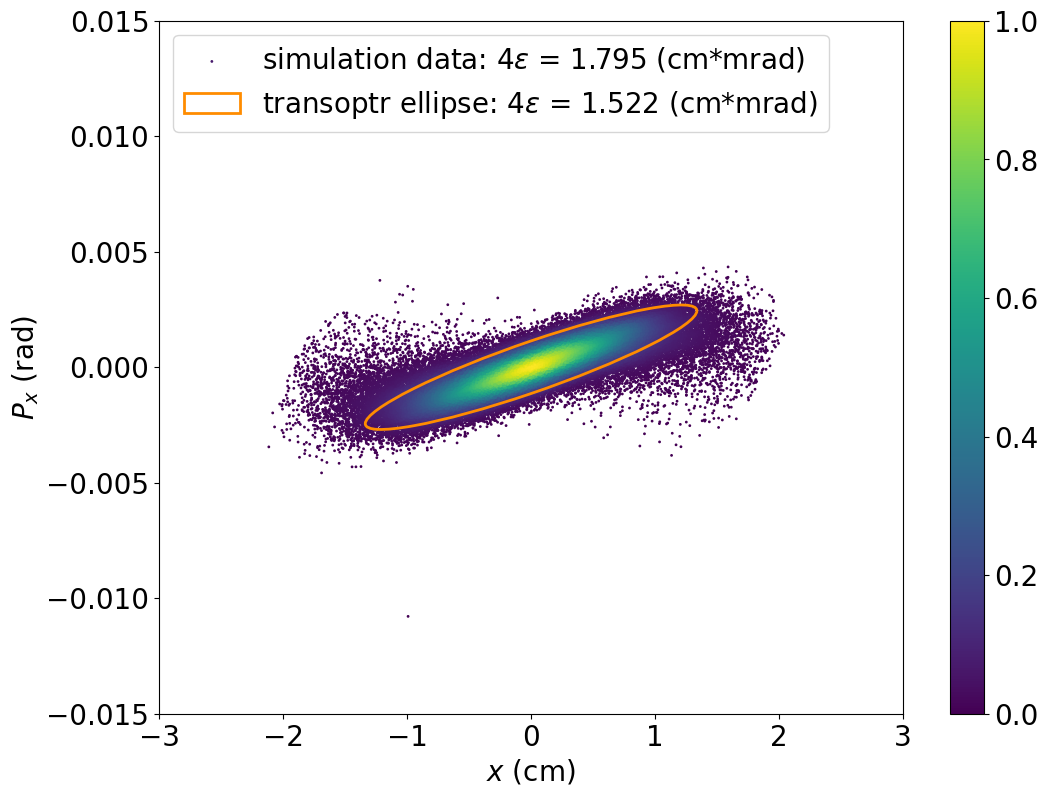}
    \includegraphics[width = 0.49\linewidth]{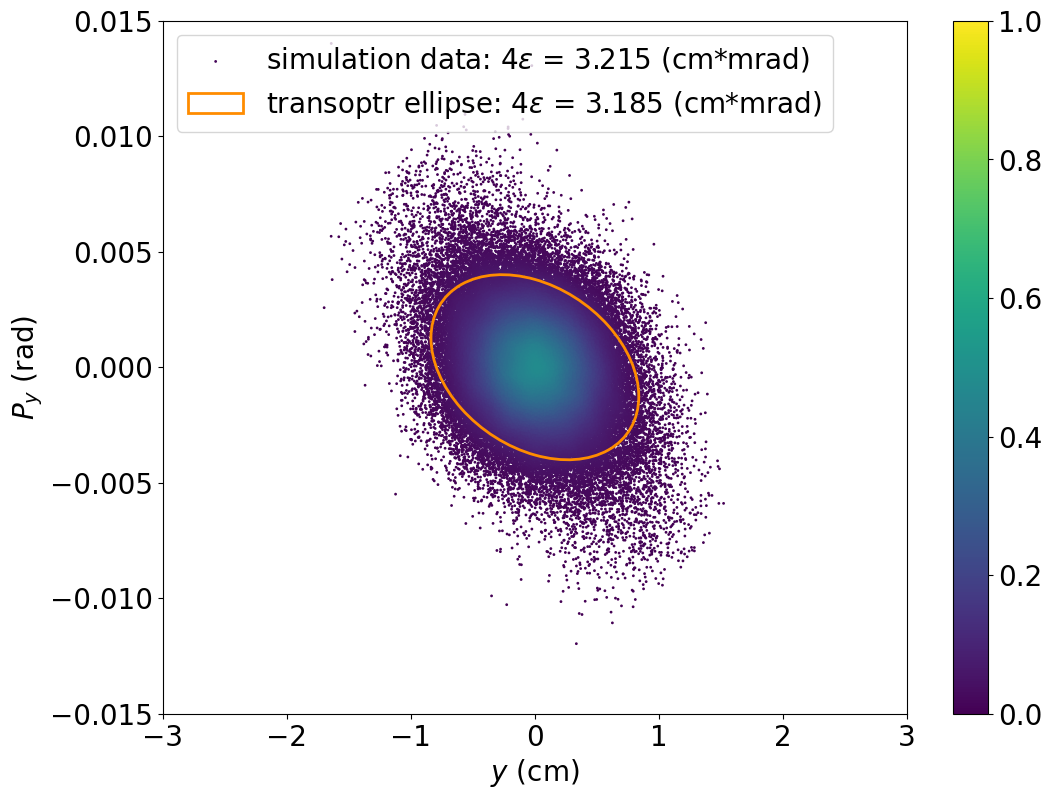}
    \caption{Phase space distributions of particles generated in FLUKA compared to the phase space ellipse obtained from {\sc transoptr} for the 5\,\SI{}{\micro\meter} target run. Top left: $x$ phase space at start of simulation; Top Right: $y$ phase space at start of simulation; Bottom left: $x$ phase space at the end of simulation; Bottom right: $y$ phase space at end of simulation. Colorbars represent the normalized density of particles.}
    \label{fig:fluka_phasespace}
\end{figure}

To compare FLUKA with {\sc transoptr}, the standard deviation of the beam particle distributions were calculated every centimeter along the beam path. The RMS of those distributions are plotted to aid in visualizing the beam envelope along the beamline.
Figure~\ref{fig:scatterFLK} presents the comparison between the beam envelopes obtained from the field parameterization input into FLUKA with the beam envelopes obtained from the code {\sc transoptr} for target thicknesses off 1.0\,\SI{}{\micro\meter}, 5.0\,\SI{}{\micro\meter}, and 10.0\,\SI{}{\micro\meter}. Figure~\ref{fig:fluka_phasespace} shows the phase space distributions in $x$ and $y$ of the 31\,MeV electrons at the start and end points of the FLUKA simulation for the 5\,\SI{}{\micro\meter} target run, overlaid with the corresponding {\sc transoptr} ellipses.

We can make similar observations from this comparison regarding the nonlinear effects as we did in Section~\ref{sec:GEANT4}. Notably, the discrepancy between envelopes increases as larger beams are transported through the optical elements, the beam emittance increases between start and end of the FLUKA simulation, and the distorted ``S" shape can again be seen in the end phase space distribution.

\subsection{Comparison of GEANT4 and FLUKA}
It is important to note that the scattering through the target is handled differently in each multi-particle code: FLUKA employs Moliere scattering theory \cite{Fontana2014FLUKA}, whereas GEANT4 employs either the Urban or Wentzel multiple scattering methods, subject to implementation \cite{kluck2026impact}. A more complete comparison between these two codes can be found in \cite{androulakaki2021comparative}. This variation in scattering methodology may be the cause of any discrepancies between these two models. 

\section{Conclusion}

As a conclusion of this work we have at our disposal a realistic, Maxwellian, three-dimensional model for the optics of soft-edge quadrupoles. The model is easily implemented into both linear optics and multi-particle codes, and can be used directly at the initial design stage, without needing the complete physical design of the element. 
There is a small cost to using this formalism in terms of computational resources, as it is necessary to numerically evaluate trigonometric functions. However, users are guaranteed to be working with realistic fields that can actually be produced by a physical magnet. 
We have demonstrated the use of these formulas to design the section of the DarkLight experiment that was successfully installed and commissioned at the TRIUMF e-Linac in Fall 2025. 

\newpage
\appendix
\section{Pseudo-code}\label{sec:appendix}

Below is an example structure of how to implement the analytic field description as laid out in Section~\ref{sec:tanh} for ease of use in the analysis code of your choice.\\

\begin{algorithmic}[1]

\Function{R}{$x, y, z, l$}
    \State \Return
    $\arctan\!\left(
        \dfrac{\sin(x)\sinh(l)}
              {\cos(x)\cosh(l) + \cosh(z)}
    \right)$
\EndFunction

\Function{S}{$x, y, z, l$}
    \State \Return
    $\dfrac{1}{2}
    \ln\!\Big(
        (\cos(x) + \cosh(l - z))
        (\cos(y) + \cosh(l + z))
    \Big)$
\EndFunction

\Function{F}{$x, y, z, l, K$}
    \State $\mathbf{v} \gets
    \begin{bmatrix}
        R(x,y,z,l) \\
        -R(y,x,z,l) \\
        S(x,y,z,l) - S(y,x,z,l)
    \end{bmatrix}$
    \State \Return $\dfrac{K}{2l}\,\mathbf{v}$
\EndFunction

\Function{E}{$x', y', z', L', K, \lambda$}
    \Comment{$L'$ is the total effective length}
    \State $x \gets \dfrac{2x'}{\lambda}$
    \State $y \gets \dfrac{2y'}{\lambda}$
    \State $z \gets \dfrac{2z'}{\lambda}$
    \State $l \gets \dfrac{L'}{\lambda}$
    \State \Return $F(x,y,z,l,K)$
\EndFunction

\Function{B}{$x', y', z', L', K, \lambda$}
    \State $\mathbf{v} \gets
    E\!\left(
        \dfrac{x' - y'}{\sqrt{2}},
        \dfrac{x' + y'}{\sqrt{2}},
        z',
        L',
        K,
        \lambda
    \right)$
    \State $\mathbf{M} \gets
    \begin{bmatrix}
        \tfrac{1}{\sqrt{2}} & -\tfrac{1}{\sqrt{2}} & 0 \\
        \tfrac{1}{\sqrt{2}} &  \tfrac{1}{\sqrt{2}} & 0 \\
        0                   &  0                   & 1
    \end{bmatrix}$
    \State \Return $\mathbf{v}\cdot\mathbf{M}$
\EndFunction

\end{algorithmic}

\begin{acknowledgments}

The authors would like to thank R.~Baartman and J.~C.~Bernauer for helpful discussions, and J. Nasser for further testing the implementation. E.~C.~is partially supported by U.S. Department of Energy Grant Number DE-FG02-94ER40818 and the U.S. National Science Foundation Grant PHY-2412703.

\end{acknowledgments}


\bibliography{References}

\end{document}